\documentclass[preprint,pre,superscriptaddress,floatfix]{revtex4}
\usepackage{amsmath,amssymb,amsthm}
\usepackage{graphicx}
\usepackage{float}

\begin{document}
\newcommand{\mt}[0]{microtubule }
\newcommand{\twod}[0]{two-dimensional }
\newcommand{\threed}[0]{three-dimensional }
\newcommand{\rmd}{\mathrm{d}}
\newcommand{\rme}{\mathrm{e}}
\newcommand{\rmi}{\mathrm{i}}

\title{Nonlinear dynamics of cilia and flagella} 
 
\author{Andreas Hilfinger}
\altaffiliation[Current address: ]{Harvard University, Department of Systems Biology, 200 Longwood Ave, Boston, MA 02115, USA}
\affiliation{Max Planck Institute for the Physics of Complex Systems, N\"othnitzer Str.~38, 01187 Dresden, Germany}

\author{Amit K Chattopadhyay}
\altaffiliation[Current address: ]{University of Delhi, Department of Physics \& Astrophysics, Delhi 110\;007, India}
\affiliation{Max Planck Institute for the Physics of Complex Systems, N\"othnitzer Str.~38, 01187 Dresden, Germany}
\affiliation{Mathematics Institute, University of Warwick, Coventry CV4 7AL, UK}

\author{Frank J\"ulicher}
\email{julicher@pks.mpg.de}
\affiliation{Max Planck Institute for the Physics of Complex Systems, N\"othnitzer Str.~38, 01187 Dresden, Germany}

\begin{abstract}
Cilia and flagella are hair-like extensions of eukaryotic cells which generate oscillatory beat patterns that can propel micro-organisms and create fluid flows near cellular surfaces. The evolutionary highly conserved core of cilia and flagella consists of a cylindrical arrangement of nine \mt doublets, called the axoneme. The axoneme is an actively bending structure whose motility results from the action of dynein motor proteins cross-linking \mt doublets and generating stresses that induce bending deformations. The periodic beat patterns are the result of a mechanical feedback that leads to self-organized bending waves along the axoneme. Using a theoretical framework  to describe planar beating motion, we derive a nonlinear wave equation that describes the fundamental Fourier mode of the axonemal beat. We study the role of nonlinearities and investigate how the amplitude of oscillations increases in the vicinity of an oscillatory instability. We furthermore present numerical solutions of the nonlinear wave equation for different boundary conditions. We find that the nonlinear waves are well approximated by the linearly unstable modes for amplitudes of beat patterns similar to those observed experimentally.
\end{abstract}

\maketitle

\section{Introduction}
Cilia and flagella are hair-like appendages of eukaryotic cells exhibiting regular, wave-like oscillations \cite{gibb81}. Their ability to generate regular beat patterns plays an important role in many systems where motion on a cellular level is required \cite{bray01}. Examples range from the propulsion of single cells, such as the swimming of sperm, to the transport of fluid along ciliated surfaces, such as the flow of mucus in the trachea. Ciliary and flagellar beat patterns are generated by an active structure called the axoneme which consists of nine microtubule doublets arranged in a cylindrical geometry \cite{gibb81,afze95,nica05}. A large number of dynein motor proteins are arranged between adjacent microtubule doublets and generate internal stresses within the axoneme that induce relative filament sliding and as a consequence axonemal bending \cite{sati65,gibb65,summ71,brok89,port00,vern02}. 

Axonemal beat patterns have been the subject of several theoretical analyses trying to elucidate the mechanisms underlying the generation of regular beat patterns \cite{mach58,brok71,brok75,brok75b,hine79,lind94a,lind94b,lind02,cama00,brok99,brok02,brok05,ried07,hilf08}. 
Recent evidence suggests that the interplay of collectively operating motors together with the elastic microtubules constitutes a mechanical feedback that leads to oscillating instabilities \cite{brok75,juli97,cama00}. The resulting traveling wave bending patterns can account for the experimentally observed beat patterns in bull sperm \cite{ried07}.

In the present article, we discuss the properties of self-organized beating patterns, extending previous work in which the linearly unstable modes near an oscillatory instability were studied \cite{cama00,ried07}. We present a nonlinear wave equation that describes the fundamental Fourier mode of planar axonemal beat patterns and derive analytically how the nonlinearities determine the amplitude of the beat beyond the bifurcation point. Furthermore, we present numerical solutions of the nonlinear wave equation subject to three different boundary conditions. 

\section{Dynamic equations of motion}
\begin{figure}[!htbp]
\begin{center}
\includegraphics[width=0.6\columnwidth]{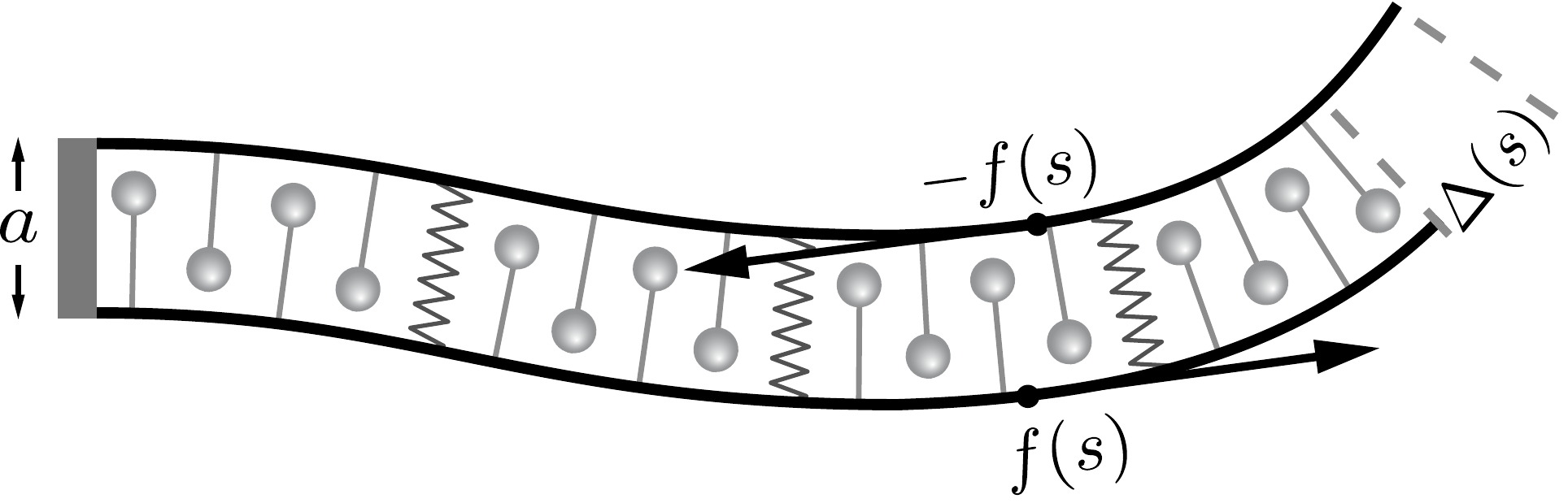}
\end{center}
\caption{
Schematic representation of the effective two-dimensional mechanics of planar beats with two elastic rods sliding relative to each other due to the shear forces generated by active elements. Illustrated are the tangential shear forces $f(s)$ and the local sliding displacement $\Delta(s)$. Elastic structural elements are indicated as springs.
}
\label{FIG: 2D axonemal projection}
\end{figure}

\begin{figure}[!htbp]
\begin{center}
\includegraphics[width=0.6\columnwidth]{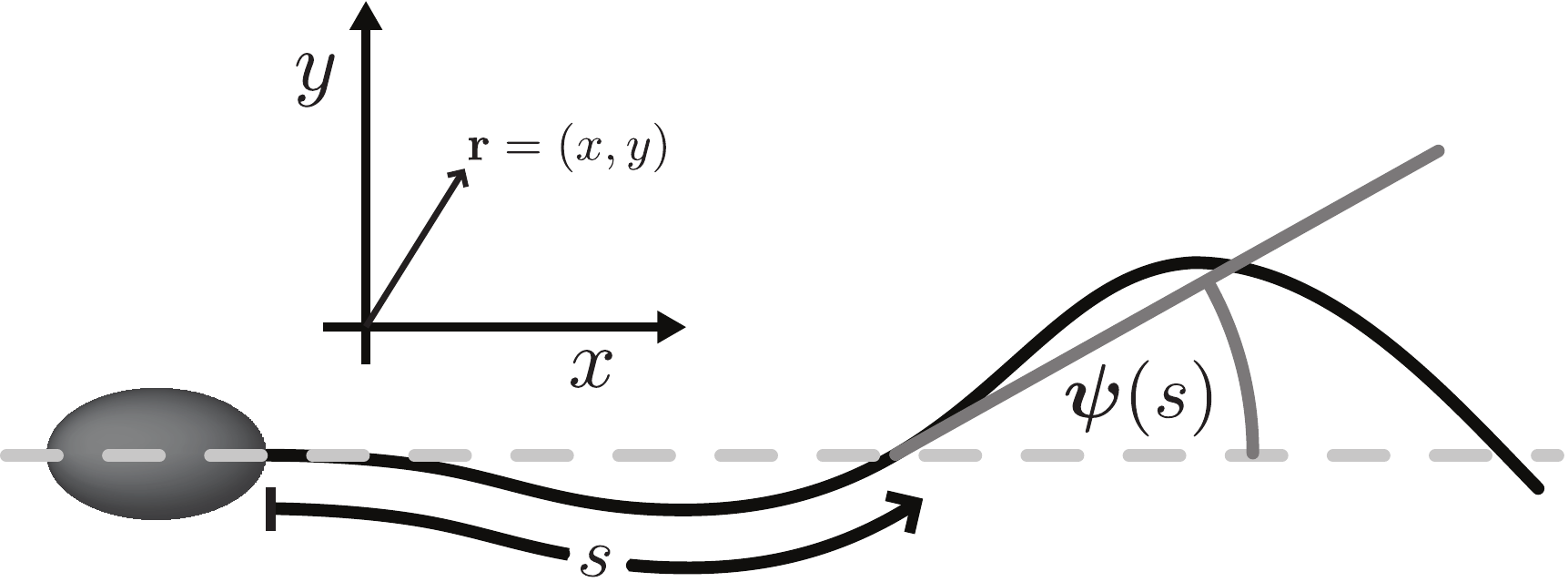}
\end{center}
\caption{Geometry of the flagellar deformation in the $x,y$-plane. The shape at a given time is described by the local tangent angle $\psi(s)$ as a function of the arc length $s$ along the flagellum.}
\label{FIG: shear angle definition}
\end{figure}

Motivated by the observation that the flagellar beat patterns of many sperm are approximately planar we discuss the dynamics of the axoneme in a plane. Such planar beat patterns can be described by an effective, two-dimensional description of the three-dimensional axonemal structure, in which the axoneme is represented by two elastic rods separated by a fixed distance $a$, corresponding to the axonemal diameter of 185 nm \cite{cama00,ried07}. These rods are linked by elastic structural elements and by active force generators, corresponding to the dynein motor proteins. To describe the relative motion of the two rods, we introduce the local sliding displacement $\Delta$ and the local shear force density $f$ exerted by passive elastic and active elements, as illustrated in figure \ref{FIG: 2D axonemal projection}.
We denote by $\mathbf{r}(s,t)$ the two-dimensional space curve parametrized by its arc length $s$ describing the shape of the centre line of the axoneme of length $L$, at time $t$. As illustrated in figure \ref{FIG: shear angle definition}, this shape can be characterized by the local tangent angle $\psi(s,t)$ such that 
\begin{equation}
\mathbf{r}(s)=\mathbf{r}(0)+\int_{0}^{s}(\cos\psi(s^{\prime}),\sin\psi(s^{\prime}))\mathrm{d}s^{\prime} \quad,
\label{EQ: 2D filament shape definition}
\end{equation}
where we have dropped the time dependence for notational convenience.
In this \twod geometry, the local sliding displacement and the local tangent angle are then related by
\begin{equation}
\Delta (s) = \Delta_0 + a \left( \psi(s) - \psi(0) \right) \quad,
\end{equation}
where $\Delta_0$ denotes the relative sliding displacement at the base \cite{vern02,vern04,ried07}. For simplicity, we ignore hydrodynamic interactions and describe the local hydrodynamic friction by introducing drag coefficients per unit length $\xi_{||}$ and $\xi_{\perp}$ for movements in directions parallel and perpendicular to the axonemal axis, respectively. The dynamics of the axoneme with a bending rigidity $\kappa$ and an internal shear force density $f(s,t)$ is then described by the following set of coupled nonlinear
equations \cite{cama00}
\begin{align}
&\partial _{t}\psi = \frac{1}{\xi _{\perp }}\left(-\kappa \psi'''' + af'' + \tau'\psi' + \tau \psi''\right) 
+ \frac{1}{\xi _{\parallel }} \left(\kappa (\psi')^{2} \psi'' - af(\psi')^{2} + \tau' \psi'\right) 
\label{EQ: psi time dependent nl dynamics}\\
&\tau'' - \frac{\xi _{\parallel }}{\xi _{\perp }}(\psi')^{2}\tau = 
a \left( f'\psi' + f\psi''\right) - \kappa \left( (\psi'')^{2} + \psi' \psi''' \right)
+ \frac{\xi _{\parallel }}{\xi _{\perp }} \left( af'\psi' - \kappa \psi'\psi''' \right) \quad,
\label{EQ: tau time dependent nl dynamics}
\end{align}
where the primes denote derivatives with respect to the arc length $s$, i.e. $\psi' \equiv \partial_s \psi$. The lateral tension $\tau(s,t)$ ensures that the filament satisfies the local inextensibility constraint $({\bf r}')^2=1$. Note that these equations can be derived from a full \threed dynamic description of the axonemal cylinder, restricted to deformations in a plane and do not require the introduction of the effective, \twod axoneme shown in figure \ref{FIG: 2D axonemal projection}.
\cite{hilf08,hilf06}. 

\section{Boundary conditions}
The dynamic equations (\ref{EQ: psi time dependent nl dynamics})--(\ref{EQ: tau time dependent nl dynamics}) are complemented by boundary conditions. While the distal end ($s=L$) is typically free to move without external constraints, the basal end ($s=0$) is subjected to external forces and torques. Matching the internal and external torques and forces at the ends determines the boundary conditions for the tangent angle $\psi(s,t)$ and the tension $\tau(s,t)$ \cite{cama00}. Different experimental conditions imply different boundary conditions at the basal end $s=0$. Motivated by experiments in which the centre of the sperm head is held at a fixed position but potentially free to pivot \cite{ried07}, we describe the dynamics of the head angle $\psi(0,t)$ by introducing an angular elastic modulus $k_{\mathrm{p}}$ and an angular friction coefficient $\gamma_{\mathrm{p}}$. The set of general boundary conditions is summarized in table \ref{TABLE: Boundary Conditions - General}. In this article we will discuss the cases of (i) a clamped head corresponding to the limit of large $k_{\mathrm{p}}$ and (ii) a freely pivoting head corresponding to $k_{\mathrm{p}}=0$ and $\gamma_{\mathrm{p}}=0$.
\begin{table}[!htbp]
\begin{tabular}{ll} \hline \hline
At $s=0$ & At $s=L$ \\ \hline
$\kappa \psi'+a\int_{0}^{L}f(s)\mathrm{d}s -k_{\mathrm{p}} \psi-\gamma_{\mathrm{p}} \partial_t\psi= 0$ & $\psi' = 0$ \\ 
$\kappa \psi''' - a\dot{f} - \psi'\tau = 0$ & $\kappa \psi'' - af = 0$ \\ 
$\kappa \psi'\psi'' - af\psi' + \tau' = 0$ & $\tau = 0$ \\ 
\hline \hline
\end{tabular}
\caption{Boundary conditions of sperm with fixed head position and free tail. We consider specifically the clamped head corresponding to the limit of large $k_{\mathrm{p}}$ and the freely pivoting head limit with $k_{\mathrm{p}}=0$ and $
\gamma_{\mathrm{p}} =0$.}
\label{TABLE: Boundary Conditions - General}
\end{table}

In order to complete the description of the basal dynamics, it is necessary to specify the mechanical properties of the basal connection which determine the relative sliding between microtubules at the base \cite{vern02,vern04}. Recently, it has been shown that such basal sliding can have an important effect on the shape of the flagellar beat \cite{ried07}. Following this previous work we characterize the visco-elastic coupling between microtubule doublets at the basal end by a basal elasticity $k_s$ and a basal friction $\gamma_s$. The basal sliding displacement $\Delta_0(t)$ then obeys \cite{ried07}
\begin{equation}
\gamma_s \partial_{t}\Delta _{0} = - k_s\Delta _{0} - \int_{0}^{L}f(s)\mathrm{d}s\quad,
\label{EQ: delta null dynamics}
\end{equation}
and in the limit for large $k_s$ and $\gamma_s$ basal sliding is suppressed. 

\section{Oscillatory dynamics}
\subsection{Fourier representation}
Time periodic beat patterns can be represented by the temporal Fourier modes $\widetilde{\psi}_n(s)$ of the tangent angle 
\begin{equation}
\psi(s,t) = \sum_{n = -\infty}^{\infty} \widetilde{\psi}_{n}(s)\rme^{\rmi n\omega t} \quad.
\end{equation}
The Fourier modes $\tilde{f}_{n}(s)$,$\widetilde{\Delta}_n(s)$ and $\tilde{\tau}_{n}(s)$ of the local shear force density $f(s,t)$, the local sliding displacement $\Delta(s,t)$ and the tension $\tau(s,t)$ are defined identically. 

The motor proteins in the axoneme generate time dependent shear forces which induce dynamic sliding displacements $\Delta(s,t)$. The relation between sliding speed and force is a collective property of the motors together with passive elements cross-linking the axoneme. This effective mechanical property of active and passive elements can be represented as  a nonlinear relation in terms of the temporal Fourier modes \cite{juli97,cama00}
\begin{equation}
\tilde{f}_1=\alpha\widetilde{\Delta}_1 + \beta\widetilde{\Delta}_1|\widetilde{\Delta}_1|^{2}+\mathcal{O}(\Delta^5)\quad.
\label{EQ: hopf response nl terms}
\end{equation}
The emergence of spontaneous oscillations is related to negative signs of the real and imaginary parts of the linear response function $\alpha$, resulting from the collective properties of many molecular motors coupled to an elastic element \cite{juli97,cama99,cama00,gril05,pecr06}. The collective effects arise from the dependence of motor transition rates on the state of the system, as for example, introduced by a load dependence of the motor detachment rate \cite{cama00,ried07}.

\subsection{Nonlinear waves} 
The self-organized nonlinear dynamics of the axoneme can then be expressed by coupled differential equations for the discrete Fourier modes of the tangent angle and the tension. The experimentally observed beat patterns of sperm are dominated by their fundamental temporal Fourier mode, with higher harmonics contributing to less than 5\% of the wave pattern \cite{ried07}. In the following, we thus neglect higher harmonics of $\psi(t)$. 

To simplify the notation in the following, we drop the tilde when referring to temporal Fourier amplitudes, defining $\psi(s) \equiv  \tilde {\psi}_1(s), \tau_0 \equiv \tilde{\tau}_{0}(s), \tau_2 \equiv \tilde{\tau}_{2}(s)$. We also introduce dimensionless parameters $\bar\omega, \bar\alpha, \bar\beta$ and $\bar{\Delta}_0$  as defined in \ref{APPENDIX: Dimensionless boundary conditions}. Taking into account nonlinearities self-consistently up to cubic terms, equations (\ref{EQ: psi time dependent nl dynamics}),(\ref{EQ: tau time dependent nl dynamics}) and (\ref{EQ: hopf response nl terms}) lead to the following set of coupled nonlinear equations for the dominant modes $\psi$, $\bar\tau_0$ and $\bar\tau_2$, as rescaled in \ref{APPENDIX: Dimensionless boundary conditions}
\begin{align}
& \rmi\bar{\omega}\psi  =-\ddddot{\psi}+\bar{\alpha}\ddot{\psi} + \bar{\beta} \partial _{\bar{s}}^{2}[(\psi +\bar{\Delta}_0-\psi(0))|\psi +\bar{\Delta}_0-\psi(0)|^{2}]+\partial_{\bar{s}}(\bar{\tau} _{0}\dot{\psi}+\bar{\tau} _{2}\dot{\psi}^{\ast }) \nonumber \\
& \hspace{3ex}+\frac{\xi_{\perp}}{\xi{\parallel}} \left[ \partial _{\bar{s}}(|\dot{\psi}|^{2}\dot{\psi})-2\bar{\alpha}(\psi +\bar{\Delta}_0-\psi(0))|\dot{\psi}|^{2}-\bar{\alpha}^{\ast }(\psi ^{\ast} + \bar{\Delta}_0^{\ast} - \psi^{\ast}(0))\dot{\psi}^{2}+\dot{\psi}\partial_{\bar{s}} \bar{\tau}_{0} + \dot{\psi}^{\ast }\partial_{\bar{s}} \bar{\tau}_{2}\right]  \nonumber \\
& \partial^2_{\bar{s}} \bar{\tau}_{0} = 2\mathrm{Re}\{\bar{\alpha}\partial _{\bar{s}}[(\psi + \bar{\Delta}_0-\psi(0))\dot{\psi}^{\ast }]\}-\partial _{\bar{s}}^{2}(|\dot{\psi}|^{2})+2\frac{\xi_{\parallel}}{\xi{\perp}}\left( |\dot{\psi}|^{2}\mathrm{Re}\left\{ \bar{\alpha}\right\} -
\mathrm{Re}\{\dot{\psi}^{\ast }\dddot{\psi}\}\right) \nonumber \\ 
& \partial^2_{\bar{s}} \bar{\tau}_{2} = \bar{\alpha}\partial _{\bar{s}}[(\psi +\bar{\Delta}_0-\psi(0))\dot{\psi}
]-\partial _{\bar{s}}(\dot{\psi}\ddot{\psi})+\frac{\xi_{\parallel}}{\xi{\perp}}\left( \bar{\alpha}
\dot{\psi}^{2}-\dot{\psi}\dddot{\psi}\right) \quad . 
\label{EQ: nl coupled equations in psi and tau}
\end{align}
Here the dots denote derivatives with respect to the rescaled arc length $\bar{s}=s/L$ and complex conjugates are denoted by asterisks. In equation (\ref{EQ: nl coupled equations in psi and tau}) we have also introduced the dimensionless linear contribution to the fundamental Fourier mode of the basal sliding displacement \cite{ried07}
\begin{equation*}
\bar{\Delta}_0 =\frac{\bar{\alpha}}{\rmi\bar{\omega}\bar{\gamma}_s+\bar{k}_s+\bar{\alpha}}
\left(\psi(0)-\int_{0}^{1}\psi (\bar{s})\mathrm{d}\bar{s} \right)\quad.
\end{equation*}
The corresponding boundary conditions complementing equation (\ref{EQ: nl coupled equations in psi and tau}) are summarized in \ref{APPENDIX: Dimensionless boundary conditions}.

\section{Wave amplitudes}
The above system exhibits an oscillatory instability or Hopf bifurcation, at which the unstable modes are described by a linearized wave equation \cite{cama00,ried07}. In the oscillatory regime close to the bifurcation, finite amplitude solutions of the full nonlinear wave equation (\ref{EQ: nl coupled equations in psi and tau}) are similar to the linearly unstable modes. The growth of the amplitude and the changes of the shape of the beating mode with increasing distance from the bifurcation are determined by the nonlinear terms of equation (\ref{EQ: nl coupled equations in psi and tau}). In the following we study the effects of nonlinearities near the bifurcation using a systematic expansion.  

Linearizing the nonlinear wave equation (\ref{EQ: nl coupled equations in psi and tau}) in the limit of small amplitudes, the linearly unstable modes denoted by $u_{0}(\bar{s})$ satisfy \cite{cama00,ried07}
\begin{equation}
{\cal L}_\mathrm{c}  u_{0}(\bar{s})=0 \quad,
\label{EQ: linear problem definition}
\end{equation}
subject to appropriate boundary conditions \cite{cama00,ried07}, 
where we have defined the linear operator 
\begin{equation}
{\cal L}(\bar\alpha,\bar\omega) = \rmi \bar\omega + \partial_{\bar{s}}^4 - \bar{\alpha} \partial_{\bar{s}}^2 \quad.
\end{equation}

\begin{figure}[!htbp]
\begin{center}
\includegraphics[width=0.55\columnwidth]{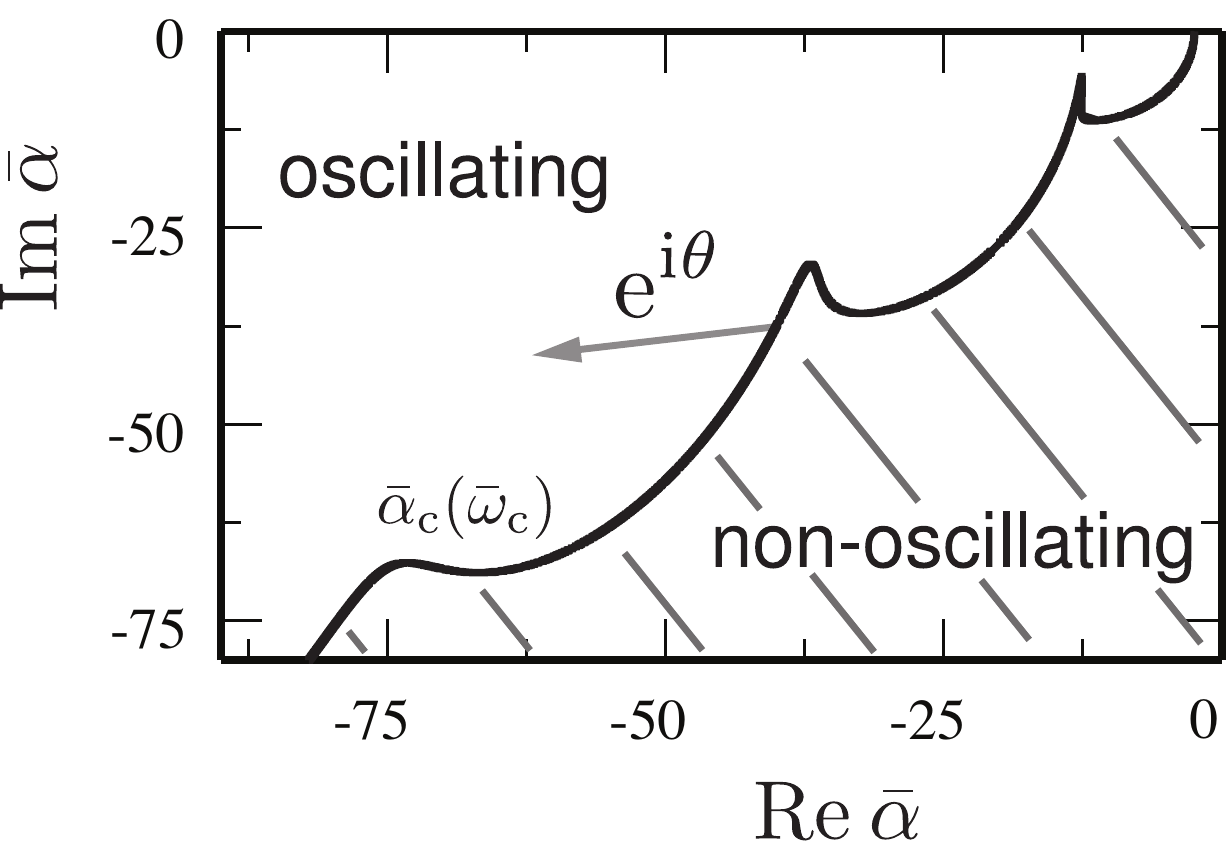}
\end{center}
\caption{Schematic diagram of the complex plane of motor impedance $\bar{\alpha}_\mathrm{c}$.  For a given dimensionless frequency $\bar\omega_\mathrm{c}$, there exists a critical value $\bar\alpha_\mathrm{c}$ describing an oscillatory instability. The line $\alpha_\mathrm{c}$ parametrized by $\bar\omega_\mathrm{c}$ is shown in black. In the dashed region to the right of the line the system is quiescent, whereas to the left of $\bar \alpha_\mathrm{c}$ it oscillates. The phase theta describes the orientation of a displacement $\delta \bar \alpha$ away from a bifurcation point in the complex plane such that $\delta \bar \alpha = |\delta \bar \alpha| e^{\rmi \theta}$.}
\label{FIG: chi diagram for going away from the instability in a given direction}
\end{figure}

Note that the amplitude of the linear mode $u_0$ is not determined by the linear equation (\ref{EQ: linear problem definition}). For convenience, we normalize $u_0$ such that $\int_{0}^{1} |u_0(\bar{s})| \rmd \bar{s}=1$. Equation (\ref{EQ: linear problem definition}), together with the appropriate boundary conditions \cite{cama00,ried07}, constitutes a boundary value problem. Nontrivial solutions exist only for pairs of critical values of the dimensionless frequency and response coefficient $(\bar\alpha_{\mathrm{c}},\bar\omega_\mathrm{c})$ \cite{cama00,hilf06}. In the following we will denote ${\cal L}_\mathrm{c}={\cal L}(\bar\alpha_\mathrm{c},\bar\omega_\mathrm{c})$, where $\bar\alpha_\mathrm{c}$ and $\bar\omega_\mathrm{c}$ are the values of $\bar\alpha$ and $\bar\omega$ at the bifurcation point.  
In figure \ref{FIG: chi diagram for going away from the instability in a given direction}, the line of critical values is indicated by the solid line representing $\bar\alpha_\mathrm{c}$ as a function of $\bar\omega_{\mathrm{c}}$. Note that there exists a discrete spectrum of such critical lines \cite{cama00}.

We can express solutions of the full nonlinear problem for parameters $\bar\alpha=\bar\alpha_\mathrm{c} + \delta\bar\alpha$ and $\bar\omega=\bar\omega_\mathrm{c}+\delta \bar\omega$ in the vicinity of the bifurcation point by an expansion of the form
\begin{align}
\psi (\bar{s}) &= \epsilon u_{0}(\bar{s}) + \epsilon ^{3}u_{1}(\bar{s}) + \mathcal{O}(\epsilon^5) 
\nonumber \\
\bar{\tau} _{0}(\bar{s}) &= \epsilon ^{2}v (\bar{s})+\mathcal{O}(\epsilon^4) 
\label{EQ: epsilon expansion of psi and tau} \\
\bar{\tau} _{2}(\bar{s})  &= \epsilon ^{2}w (\bar{s})+\mathcal{O}(\epsilon^4) 
\nonumber \quad.
\end{align}

Here $\epsilon$ is a small dimensionless number that characterizes the distance from the bifurcation point  by
\begin{equation}
\delta \bar{\alpha} =\rho \rme^{\rmi\theta }\epsilon ^{2} \hspace{2.0ex} \mathrm{and} \hspace{2.0ex} \delta \bar{\omega} =\mu \epsilon ^{2}  \quad ,
\label{EQ: delta_chi and delta_omega epsilon expansion}
\end{equation}
where we have introduced the real coefficients $\rho$ and $\mu$ as well as the phase $\theta$.

Inserting the ansatz (\ref{EQ: epsilon expansion of psi and tau}) into the wave equation (\ref{EQ: nl coupled equations in psi and tau}), we can solve this equation systematically order by order near a given bifurcation point. First, the linearly unstable mode $u_0(\bar s)$ is determined. Then with $u_0(\bar s)$ known, we can determine the static and dynamic tension profiles $v(\bar{s})$ and $w(\bar{s})$, in terms of $u_{0}(\bar{s})$ as detailed in \ref{APPENDIX: First and second order terms}.

Matching terms to third order in $\epsilon$ then leads to an equation for the nonlinear correction  $u_{1}(\bar{s})$
to the waveform
\begin{equation}
\mathcal{L}_{\mathrm{c}}u_{1}=\rho \rme^{\rmi \theta }\partial _{\bar{s}}^{2}u_{0} - \rmi \mu u_{0}
- \mathcal{N} (u_{0})  \quad,
\label{EQ: ampl determining equ}
\end{equation}
where the nonlinear terms $\mathcal{N}(u_{0})$ are given by
\begin{align}
\mathcal{N}(u_{0}) =& 
\bar{\beta}\partial _{\bar{s}}^{2} \left[(u_{0} +\bar{\Delta}_0^{(\mathrm{c})}-u_{0}(0))|u_{0} +\bar{\Delta}_0^{(\mathrm{c})}-u_{0}(0)|^{2}\right] 
+ \partial_{\bar{s}}\left(v\dot{u}_{0}+w \dot{u}_{0}^{\ast } \right) \nonumber \\
&+ \frac{\xi_{\perp}}{\xi{\parallel}} \left[ \partial _{\bar{s}} (|\dot{u}_{0}|^{2}\dot{u}_{0}) 
- 2\bar{\alpha} (u_{0} +\bar{\Delta}_0^{(\mathrm{c})}-u_{0}(0))|\dot{u}_{0}|^{2} \right . \nonumber \\
&\hspace{6ex}\left .-\bar{\alpha}^{\ast } \left(u_{0}^{\ast
} +\bar{\Delta}_0^{(\mathrm{c})}-u_{0}^{\ast}(0)\right) (\dot{u}_{0})^{2}+\dot{v}\dot{u}_{0}+\dot{w}
\dot{u}_{0}^{\ast }\right]  \quad,
\label{EQ: Definition of NL-terms in perturbation equation}
\end{align}
and $\bar{\Delta}_0^{(\mathrm{c})}$ is the basal sliding term evaluated at the bifurcation point, as defined by equation (\ref{EQ: basal sliding to linear order}) in the Appendix. 

Using equation (\ref{EQ: ampl determining equ}) we can obtain a relation between the coefficients $\rho$,$\mu$ and $\theta$ without calculating the nonlinear correction $u_1$. This is achieved by introducing a function $u_{0}^{+}$ adjunct to the linear mode $u_0$ which has the property $\mathcal{L}_{\mathrm{c}}u_{0}^{+} = 0$ and obeys $\int_{0}^{1}u_{0}^{+}\mathcal{L}_{\mathrm{c}}u_1=Z$. The constant $Z$ depends on $\theta,\rho,\mu$ and is derived explicitly in \ref{APPENDIX: Third order terms} for three different mechanical conditions imposed at the basal end. Multiplication of (\ref{EQ: ampl determining equ}) with $u_{0}^{+}$ and subsequent integration leads to 
\begin{equation}
- \rho \rme^{\rmi \theta }\int_{0}^{1}u_{0}^{+}\partial _{\bar{s}}^{2}u_{0}\mathrm{d}\bar{s}
+ \rmi \mu \int_{0}^{1}u_{0}^{+}u_{0}\mathrm{d}\bar{s}
+ \int_{0}^{1}u_{0}^{+}\mathcal{N}(|u_{0}|^{2}u_{0})\mathrm{d}\bar{s}+Z=0 \hspace{1ex}.
\label{EQ: nl amplitude equation}
\end{equation}

We can now discuss the emergence of the unstable mode $\psi (\bar{s})$ and its frequency $\bar \omega$ when starting at a bifurcation point at $\bar \alpha=\bar \alpha_\mathrm{c}$ as illustrated in figure \ref{FIG: chi diagram for going away from the instability in a given direction}: moving from $\bar \alpha_\mathrm{c}$ in a direction given by an angle $\theta$, equation (\ref{EQ: nl amplitude equation}) describes the beating mode in the oscillatory region of the state diagram characterized by $\bar \alpha = \bar \alpha_\mathrm{c}+ |\delta \bar \alpha| e^{\rmi \theta}$. 
For a chosen value of $\theta$, the values of $\rho$ and $\mu$ can be uniquely determined from the complex equation (\ref{EQ: nl amplitude equation}). Equation (\ref{EQ: delta_chi and delta_omega epsilon expansion}) then describes the increase of the amplitude as
\begin{equation}
\epsilon = \left( \frac{|\delta \bar{\alpha}|}{\rho} \right)^{1/2}
\label{EQ: square root growth of epsilon}
\end{equation}
while the frequency changes at the same time by
\begin{equation}
\delta \bar{\omega} = \frac{\mu}{\rho}|\delta \bar{\alpha}| \quad .
\end{equation}

Note that these behaviors depend on the angle $\theta$ chosen. Two special situations are of interest. There exists in general a specific choice $\bar \theta$ for which $\mu=0$, i.e. in this case the frequency does not change when moving in the corresponding direction away from the bifurcation line. Examples for such lines in the complex plane along which the frequency of the unstable modes remains the same as at the bifurcation point $\bar \alpha_\mathrm{c}$ are displayed in figure \ref{FIG: numerical nonlinear solutions}. A second special choice 
$\theta = \theta_\parallel$ is the direction tangential to the bifurcation line $\bar \alpha_\mathrm{c}(\bar \omega_\mathrm{c})$. For this choice, equation (\ref{EQ: nl amplitude equation}) becomes singular with $\rho \to \infty$ and $\rho/\mu=\left|\rmd \bar \alpha_\mathrm{c}/\rmd \bar \omega_\mathrm{c}\right|$. In this case the amplitude $\epsilon$ remains zero, but the frequency changes along the bifurcation line \cite{cama00}. Note also, that the shape and frequency of the beating mode at a point $\alpha$ do not depend on the reference bifurcation point from which it is reached.

The above method permits us to calculate the amplitude and frequency of beating modes close to the bifurcation line. In order to study the influence of nonlinearities on the shapes of these modes, we make use of the above analytical result to solve the nonlinear equations numerically.

\section{Numerical solutions to the nonlinear wave equations}
Periodic and planar beating patterns are solutions to the nonlinear wave equation (\ref{EQ: nl coupled equations in psi and tau}) together with the boundary conditions given in table \ref{TABLE: Fundamental modes boundary conditions} which constitute a boundary value problem, that can be solved numerically by a shooting and matching procedure \cite{pres02}. Note that the wave equation is invariant with respect to changes in the overall phase of $\psi(\bar{s})$. To remove this degeneracy we impose an arbitrary condition for the phase of $\dot{\psi}(\bar{s})$ at the base $\bar s = 0$.

In order to obtain numerical solutions that satisfy the wave equation with given boundary conditions, we first determine an approximate solution \mbox{$\psi(\bar s) \approx \epsilon u_0(\bar s)$},\mbox{$\bar{\tau} _{0}(\bar{s}) \approx \epsilon ^{2}v (\bar s)$},$\bar{\tau} _{2}(\bar{s})  \approx \epsilon ^{2}w (\bar s)$ close to a bifurcation point $\bar \alpha_\mathrm{c},\bar \omega_\mathrm{c}$ making use of the method discussed in the preceding section. This allows us to determine the amplitude $\epsilon$ and the functions $v (\bar s), w (\bar s)$ for a solution close to the bifurcation point $\bar \alpha_\mathrm{c},\bar \omega_\mathrm{c}$. This approximate solution is used as seed for the shooting and matching procedure to solve equation (\ref{EQ: nl coupled equations in psi and tau}).
 
\begin{figure}[!htbp]
\begin{center}
\includegraphics[width=0.9\columnwidth]{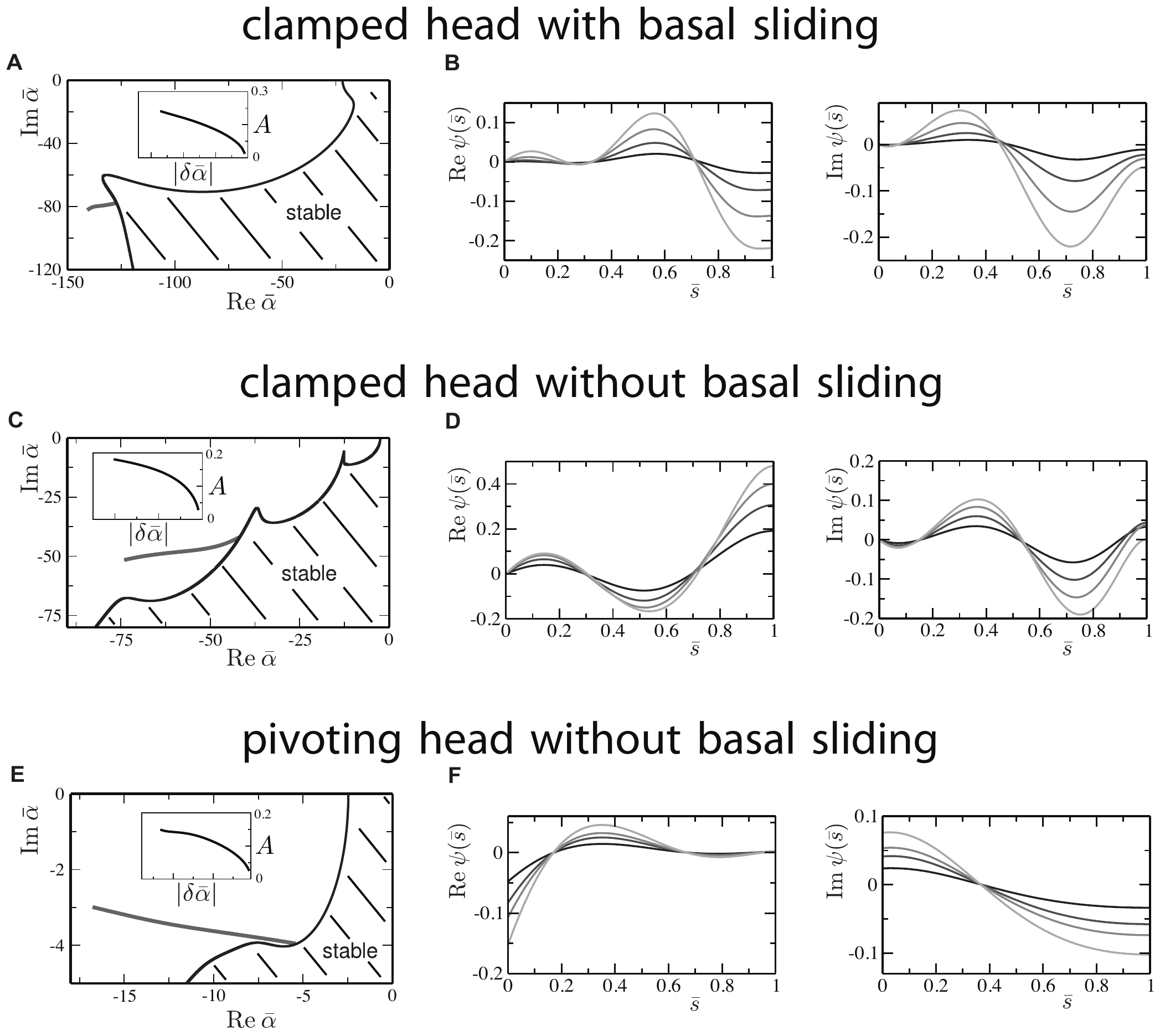}
\end{center}
\caption{Examples of beating modes, which are solutions to the wave equation (\ref{EQ: nl coupled equations in psi and tau}) for different boundary conditions. The location of these solutions is indicated in the complex plane of values of the linear response function of motors $\bar\alpha$. The chosen examples correspond to a situation where starting from a Hopf bifurcation, the amplitude of the unstable mode increases when moving along a path in the $\bar \alpha$ plane for which the oscillation frequency is  constant. (A) Solutions for clamped head with basal sliding 
with frequency $\omega / 2\pi \simeq 26$ Hz in the $\bar \alpha$ plane (grey line). The parameters $k_s$ and $\gamma_s$ were chosen as determined in \cite{ried07} such that the beat patterns shown in (B) resemble experimentally observed ones. The amplitude $A$ as a function of $\vert \delta\bar \alpha\vert$ is displayed in the inset. (B) Real and imaginary part of the waveform  $\psi$ as a function of dimensionless arc length $\bar s$ for different points along the grey line in (A). (C,D) Same as (A,B) but for the clamped head boundary conditions without basal sliding and frequency $\simeq28$ Hz. 
Here we choose the bifurcation on the first branch of unstable modes. (E,F) Same as (C,D) but for a freely pivoting head without basal sliding and frequency $\simeq 5$ Hz.
}
\label{FIG: numerical nonlinear solutions}
\end{figure}

With this procedure, we generate a sequence of numerical solutions starting from the bifurcation line and moving in a direction in which the frequency $\bar \omega$ remains constant. Examples of these solutions to equation (\ref{EQ: nl coupled equations in psi and tau}) are displayed in figure \ref{FIG: numerical nonlinear solutions} for different boundary conditions and starting from different bifurcation points.

Parameter values used in these calculations are $\xi_{\perp} = 3.4 \times 10^{-3}\mathrm{N \cdot s \cdot m^{-2}}$, $\xi_{\parallel} = \xi_{\perp} / 2$, $\kappa = 1.7 \times 10^{-21}\; \mathrm{N\cdot m}^2$, $L = 58.3$ $\mu$m as estimated for bull sperm flagella \cite{ried07,howa01}. The value of $\bar \beta$ has not been measured. We choose $\bar \beta = 42$  for which the bifurcation is supercritical.

The region of stability of the non-oscillating state is indicated in figure \ref{FIG: numerical nonlinear solutions}(A,C,E). The oscillatory instability occurs along the solid black line. The real and imaginary parts of $\psi$ for these solutions are displayed in figure \ref{FIG: numerical nonlinear solutions}(B,D,F) as a function of the dimensionless arc length $\bar{s}$  for distinct values of $\alpha$, located along the grey line shown in figure \ref{FIG: numerical nonlinear solutions}(A,C,E). As indicated by the insets, the amplitude $A = \int_{0}^{1} |\psi(\bar{s})| \rmd \bar{s}$ of the modes grows continuously with increasing distance from the instability. In the limit of small amplitudes, it obeys equation (\ref{EQ: square root growth of epsilon}) with $\epsilon\simeq A$.

In contrast to the amplitude, the shape of the beat patterns changes only weakly as illustrated by figure \ref{FIG: rescaled nonlinear solutions} in the Appendix. Solutions to the linearized equations therefore provide good approximations to the full nonlinear problem in the range of examined parameters.

\section{Conclusions and outlook}
The beating of cilia and flagella in viscous media is the result of the co-operative action of many dynein motors interacting with the elastic microtubules within the axoneme. The solutions to the linearized dynamic equations \cite{cama00} have been shown to be good approximations for experimentally observed planar beat patterns of bull sperm \cite{ried07}. However, the flagellar beat is a fundamentally nonlinear problem and cannot be satisfactorily described by a linear theory.

Here, we have presented a theory of flagellar beat patterns taking into account the leading nonlinearity. We derived the nonlinear wave equations for the fundamental modes of the beat shape and the lateral tension (\ref{EQ: nl coupled equations in psi and tau}). Furthermore, by solving the nonlinear wave equation numerically we explored how the shape of beat patterns changes as the beat amplitude increases. We found that for the parameter values and amplitude range examined here, the shape of the beat patterns of finite amplitude remain qualitatively similar to the linear waveforms obtained near the instability in the limit of infinitesimally small amplitudes. Experimentally observed wave forms of beating sperm can therefore be approximated by linear modes of \emph{ad hoc} amplitude \cite{ried07}. 
Existing nonlinear discussions of slender rod dynamics in the viscous regime have focussed on the dynamics of passive rods \cite{wigg98b,wigg98,yu06}, or described actively beating filaments subject to pre-described internal force distributions \cite{fu08}. Our analytical results provide the framework to discuss the lowest order nonlinear effects of self-organized axonemal beat patterns allowing us to bridge the gap between the linear and nonlinear regime without having to resort to a full simulation of the system.

Our results are independent of the specific molecular mechanisms underlying the collective action of motors. However, our theory does therefore not allow us to predict the influence of experimentally controlled parameters such as ATP concentration or temperature on the beat shape. To understand the influence of such parameters on the beat, more specific models of motor action are required. The oscillation frequency of the system is selected via the self-organization of dynein motors and microtubules. This frequency selection involves the full frequency dependent impedance of motors and thus also depends on details of the underlying mechanisms \cite{cama00}. A key open challenge for the understanding of flagellar dynamics is therefore to understand the parameters which govern frequency selection based on the properties of dynein motors in the system. Also, there exist several branches in parameter space along which linear instabilities occur. Which of these branches becomes unstable first and governs the behavior of the flagellar beat also depends on molecular details and will be an important challenge for future work.

\begin{acknowledgments}
We thank K.~Kruse, A.~Vilfan, I.~Riedel-Kruse and J.~Howard for many helpful discussions. AKC acknowledges Marie Curie Incoming International Fellowship number MIFI-CT-008608 for partial research support and AH acknowledges a postdoctoral fellowship by the DAAD (Deutscher Akademischer Austauschdienst) for partial research support. 
\end{acknowledgments}

\appendix
\section{Dimensionless parameters and boundary conditions}
\label{APPENDIX: Dimensionless boundary conditions}
\begin{table}[!htbp]
\begin{tabular}{lllll} \hline \hline
$\bar{\Delta}_0 = \Delta_0 / a$ & $\bar{\omega}=\frac{\omega L^{4}}{\kappa }\xi _{\perp }$ & $\bar{\alpha}=\frac{a^{2}L^{2}}{\kappa }\alpha$ 
& $\bar{\beta}=\frac{a^{4}L^{2}}{\kappa }\beta$ & $\bar{s} = \frac{s}{L}$ \\ 
$\bar{\tau} _{i} (\bar{s})=\frac{L^{2}}{\kappa }\tau _{i}(s)$ & $\bar{k}_s=\frac{a^{2}L}{\kappa }k_s$ & $\bar{\gamma}_s=\frac{a^{2}}{L^{3}\xi
_{\perp }}\gamma_s $ & $\bar{k}_{\mathrm{p}}=\frac{L}{\kappa}k_{\mathrm{p}}$ & $\bar{\gamma}_{\mathrm{p}}=\frac{a^{2}}{L^{3}\xi_{\perp }}\gamma_{\mathrm{p}}$\\ 
\hline 
\end{tabular}
\caption{Summary of the relations between the physical parameters and the dimensionless quantities.}
\label{TABLE: Dimensionless parameters}
\end{table}
Table \ref{TABLE: Dimensionless parameters} provides the definitions of dimensionless parameters used in this work. The mechanical conditions at the boundary (see table \ref{TABLE: Boundary Conditions - General}) impose 
boundary conditions for the fundamental modes $\psi(\bar{s})$, $\bar{\tau}_0(\bar{s})$ and $\bar{\tau}_2(\bar{s})$ as summarized in table \ref{TABLE: Fundamental modes boundary conditions}, 
where we have introduced the nonlinear contribution to the dimensionless basal sliding displacement
\begin{equation*}
\bar{\Delta}_0^{\mathrm{(nl)}} = \bar{\Delta}_0 - \frac{\bar{\beta}}{\rmi\bar{\omega}
\bar{\gamma}_s+\bar{k}_s + \bar{\alpha}} \int_{0}^{1}|\psi (\bar{s}) - \psi(0) + \bar{\Delta}_0|^{2}(\psi (\bar{s}) - \psi(0) +\bar{\Delta}_0)\mathrm{d}\bar{s} \quad .
\end{equation*}

\begin{table}[!htbp]
\begin{align*}
&(\bar{k}_{\mathrm{p}}+\rmi\bar{\omega}\bar{\gamma}_{\mathrm{p}})\psi(0) = \dot{\psi}(0)+\bar{\alpha}\int_{0}^{1}\left[\psi (\bar{s})+\bar{\Delta}_0-\psi (0)\right]\mathrm{d}\bar{s}\\
&\hspace{18ex}+\bar{\beta}\int_{0}^{1}\left[|\psi (\bar{s})+\bar{\Delta}_0-\psi (0)|^{2}(\psi (\bar{s})+\bar{\Delta}_0-\psi (0))\right]\mathrm{d}\bar{s}\\
&\dddot{\psi}(0) = \bar{\alpha}\dot{\psi}(0)+\dot{\psi}(0)\bar{\tau} _{0}(0)+\dot{\psi}^{\ast }(0)\bar{\tau} _{2}(0)
+\bar{\beta}(2|\bar{\Delta}_0|^{2}\dot{\psi}(0)+(\bar{\Delta}_0)^{2}\dot{\psi}^{\ast }(0))\\
&\left. \frac{\partial \bar{\tau}_{0}}{\partial \bar{s}}\right|_{\bar{s}=0} = -\partial_{\bar{s}}(|\dot{\psi}(0)|^{2})+2\mathrm{Re} \big\{\bar{\alpha}\bar{\Delta}_0\dot{\psi}^{\ast }(0)\big\}  \\
&\left. \frac{\partial \bar{\tau}_{2}}{\partial \bar{s}}\right|_{\bar{s}=0} = -\dot{\psi}(0)\ddot{\psi}(0)+\bar{\alpha}\bar{\Delta}_0\dot{\psi}(0)\\
&\dot{\psi}(1) = 0\\
&\ddot{\psi}(1) = \bar{\alpha}(\psi (1)+\bar{\Delta}_0^{\mathrm{(nl)}}+\psi (0))+\bar{\beta}|\psi(1)+\bar{\Delta}_0-\psi (0)|^{2}(\psi (1)+\bar{\Delta}_0-\psi (0))\\
&\bar{\tau} _{0}(1) = 0\\
&\bar{\tau} _{2}(1) = 0
\end{align*}
\caption{The boundary conditions for the fundamental modes $\psi(\bar{s})$, $\bar{\tau}_0(\bar{s})$ and $\bar{\tau}_2(\bar{s})$.}
\label{TABLE: Fundamental modes boundary conditions}
\end{table}
Note that for the limiting cases under consideration in the main text, 
the boundary conditions simplify as follows. In the general clamped head case we have $\psi(0)=0$, in the absence of basal sliding we furthermore have $\bar{\Delta}_0=0$, and $\bar{\Delta}_0^{\mathrm{(nl)}}=0$. For a freely pivoting head without basal sliding, $k_{\mathrm{p}}=0$, $\gamma_{\mathrm{p}}=0$, 
$\bar{\Delta}_0=0$, and $\bar{\Delta}_0^{\mathrm{(nl)}}=0$.

\section{Nonlinear perturbation calculation to second order}
\label{APPENDIX: First and second order terms}
Formally expanding the linear operator ${\mathcal{L}}$ close to the bifurcation point, we write  
\begin{equation}
\mathcal{L}(\bar{\alpha},\bar{\omega}) = \mathcal{L}_{\mathrm{c}}-\delta \bar{\alpha}\partial_{\bar{s}}^{2}+\rmi\delta \bar{\omega}\quad.
\end{equation}
Substituting the ansatz of (\ref{EQ: epsilon expansion of psi and tau}) and (\ref{EQ: delta_chi and delta_omega epsilon expansion}) into the nonlinear wave equation (\ref{EQ: nl coupled equations in psi and tau}) reproduces to linear order the equation describing the linearly unstable modes (\ref{EQ: linear problem definition}) supplemented by the appropriate boundary conditions (explicitly described in \cite{hilf06}).

Matching terms to second order in $\epsilon$ then leads to
\begin{align}
\ddot{v}(\bar{s}) &= 2\mathrm{Re}\{\bar{\alpha}_{\mathrm{c}}\partial _{\bar{s}}[(u_{0} + \bar{\Delta}_0^{(\mathrm{c})}-u_{0}(0))\dot{u}_{0}^{\ast }]\}-\partial _{\bar{s}}^{2}(|\dot{u}_{0}|^{2})
+2\frac{\xi _{\parallel }}{\xi _{\perp }}\left( |\dot{u}_{0}|^{2}\mathrm{Re}\left\{ \bar{\alpha}_{\mathrm{c}}\right\} -
\mathrm{Re}\{\dot{u}_{0}^{\ast }\dddot{u}_{0}\}\right) \nonumber \\
\ddot{w}(\bar{s}) &= \bar{\alpha}_{\mathrm{c}}\partial _{\bar{s}}[(u_{0} +\bar{\Delta}_0^{(\mathrm{c})}-u_{0}(0))\dot{u}_{0}
]-\partial _{\bar{s}}(\dot{u}_{0}\ddot{u}_{0})+\frac{\xi _{\parallel }}{\xi _{\perp }}\left( \bar{\alpha}_{\mathrm{c}}
(\dot{u}_{0})^{2}-\dot{u}_{0}\dddot{u}_{0}\right) 
\label{EQ: tau_0 and tau_2 nl correction equation}
\end{align}
where 
\begin{equation}
\bar{\Delta}_0^{(\mathrm{c})}=-\frac{\bar{\alpha}_{\mathrm{c}}}{\rmi\bar{\omega}_{\mathrm{c}}\bar{\gamma}_s+\bar{k}_s + \bar{\alpha}_{\mathrm{c}}}\int_{0}^{1}u_{0}(\bar{s})\mathrm{d}\bar{s} \quad,
\label{EQ: basal sliding to linear order}
\end{equation}
is the amplitude of basal sliding to linear order. The above system of equations, together with the appropriate boundary conditions \cite{hilf06}, allows us to obtain $v(\bar{s}),w(\bar{s})$ for given solutions $u_{0}(\bar{s})$ of the linear problem.

\section{Nonlinear perturbation calculation to third order}
\label{APPENDIX: Third order terms}
We define a function $u_0^+$ adjunct to $u_0$ such that the integral
\begin{align*}
\int_{0}^{1}u_{0}^{+}\mathcal{L}_{\mathrm{c}}u_{1} =&\int_{0}^{1}u_{0}^{+}(\rmi\bar{\omega}_{\mathrm{c}}u_{1}+\ddddot{u}_{1}-\bar{\alpha}_{\mathrm{c}}\ddot{u}_{1}) \\
=&\int_{0}^{1}u_{1}(\rmi\omega _{\mathrm{c}}u_{0}^{+}+\ddddot{u}_{0}^{+}-\bar{\alpha}_{\mathrm{c}}\ddot{u}_{0}^{+}) \\
&+\left[ u_{0}^{+}\dddot{u}_{1}-
\dot{u}_{0}^{+}\ddot{u}_{1}+(\ddot{u}_{0}^{+}-\bar{\alpha}_{\mathrm{c}}u_{0}^{+})\dot{u}_{1}-(\dddot{u}_{0}^{+}-\bar{\alpha}_{\mathrm{c}}\dot{u}_{0}^{+})u_{1}\right] _{\bar{s}=0}^{\bar{s}=1}\quad.
\end{align*}
is independent of $u_1$.
In order to eliminate $u_{1}(\bar{s})$ from the bulk term in the above expression we require that $\mathcal{L}_{\mathrm{c}}u_{0}^{+}(\bar{s})=0$, implying that $u_{0}^{+}(\bar{s})$ satisfies the same differential equation as the linear modes $u_{0}(\bar{s})$. The condition that terms depending on $u_{1}$ vanish at the boundaries $\bar{s}=0$ and $\bar{s}=1$, then defines the boundary conditions for $u_0^+$. The so-defined function $u_0^+$ then leads to $\int_0^1 u_0^+{\mathcal L}_c u_1 =Z$, where the complex constant $Z=Z(\rho,\mu,\theta)$ does not depend on the unknown correction function $u_1$. This complex constant characterizes the behavior of the system away from the bifurcation in the direction of $\theta$ and can be determined from the linearly unstable mode $u_0$ and its adjunct function $u_{0}^{+}$ only. In the following we state explicitly the boundary conditions specifying the adjunct function and determine the constant $Z$ for the various mechanical conditions of interest (clamped head without basal sliding, clamped head with basal sliding, freely pivoting head without basal sliding). These results were obtained by matching terms of $\mathcal{O}(\epsilon )$--$\mathcal{O}(\epsilon^3)$ of the perturbation ansatz (\ref{EQ: epsilon expansion of psi and tau}) substituted into the nonlinear wave equation (\ref{EQ: nl coupled equations in psi and tau}).

\subsection{Clamped head without basal sliding}
The boundary conditions for the conjugate linear solutions $u_{0}^{+}(\bar{s})$ are given by
\begin{align}
\dot{u}_{0}^{+}(0) = 0, \quad \ddot{u}_{0}^{+}(0) = 0, \quad & u_{0}^{+}(1) = 0, \quad \dddot{u}_{0}^{+}(1) = 0 \quad,
\label{EQ: psi_conj_bc_clamped_wo_bsliding}
\end{align}
which leads to 
\begin{equation}
\int_{0}^{1}u_{0}^{+}\mathcal{L}_{\mathrm{c}}u_{1} = -u_{0}^{+}(0)A -\dot{u}_{0}^{+}(1) B \nonumber \equiv Z \quad,
\label{EQ: Z equation - clamped head no basal sliding}
\end{equation}
where we have introduced
\begin{align*}
A&=\rho \rme^{\rmi\theta}\dot{u}_{0}(0) + v(0) \dot{u}_{0}(0) + w(0) \dot{u}^{*}_{0}(0) \\
B&=\rho \rme^{\rmi\theta}u_{0}(1) + \bar{\beta}\left|u_{0}(1)\right|^{2}u_{0}(1)\quad.
\end{align*}

\subsection{Clamped head with basal sliding}
The boundary conditions for the conjugate linear solutions $u_{0}^{+}(\bar{s})$ are given by
\begin{align}
\dot{u}_{0}^{+}(0)=0, \quad \ddot{u}_{0}^{+}(0)=0, \quad 
u_{0}^{+}(1) = \frac{\bar{\alpha}_{\mathrm{c}}^{2}}{\rmi\bar{\omega}_{\mathrm{c}}}\frac{1}{\rmi\bar{\omega}_{\mathrm{c}}\bar{\gamma}_s+\bar{k}_s+\bar{\alpha}_{\mathrm{c}}}\dot{u}_{0}^{+}(1), \quad \dddot{u}_{0}^{+}(1)=0, \nonumber \\
\label{EQ: psi_conj_bc_clamped_with_bsliding}
\end{align}
which leads to $Z$ of the same form as (\ref{EQ: Z equation - clamped head no basal sliding}), but with a more complicated expression for $A$ and $B$ which are now given by
\begin{align*}
A =& \rho \rme^{\rmi\theta }\dot{u}_{0}(0) + \bar{\beta}(2|\bar{\Delta}_0^{(\mathrm{c})}|^{2}\dot{u}_{0}(0) + (\bar{\Delta}_0^{(\mathrm{c})})^{2}\dot{u}_{0}^{\ast }(0)) + v(0) \dot{u}_{0}(0)+w(0) \dot{u}_{0}^{\ast }(0) \\ 
B =& \rho \rme^{\rmi\theta }(u_{0}(1) + \bar{\Delta}_0^{(\mathrm{c})}) + \bar{\beta}\left\vert u_{0}(1) + \bar{\Delta}_0^{(\mathrm{c})}\right\vert ^{2}u_{0}(1) - \frac{\mu \bar{\Delta}_0^{(\mathrm{c})}}{\bar{\omega}_{\mathrm{c}}}\bar{\alpha}_{\mathrm{c}} \\
&+ \frac{\bar{\alpha}_{\mathrm{c}}^{2}}{\rmi\bar{\omega}_{\mathrm{c}}\bar{\gamma}_s + \bar{k}_s + \bar{\alpha}_{\mathrm{c}}}\frac{1}{\rmi\bar{\omega}_{\mathrm{c}}} \Big(\rho \rme^{\rmi\theta }\dot{u}_{0}(0)-A + \int_{0}^{1}\mathcal{N}(|u_{0}(\bar{s})|^{2}u_{0}(\bar{s}))\mathrm{d}\bar{s}\Big) \\ 
&+ \frac{1}{\rmi\bar{\omega}_{\mathrm{c}}\bar{\gamma}_s+\bar{k}_s+\bar{\alpha}_{\mathrm{c}}}\Big(\bar{\Delta}_0^{(\mathrm{c})}(\rho \rme^{\rmi\theta }(\rmi\bar{\omega}_{\mathrm{c}}\bar{\gamma}_s+\bar{k}_s) 
- \rmi\mu \bar{\gamma}_s\bar{\alpha}_{\mathrm{c}})-\bar{\beta}\bar{\alpha}_{\mathrm{c}}\int_{0}^{1}|u_{0}(\bar{s})+\bar{\Delta}_0^{(\mathrm{c})}|^{2}(u_{0}(\bar{s})+\bar{\Delta}_0^{(\mathrm{c})})\mathrm{d}\bar{s}\Big) \quad,
\end{align*}
with $\mathcal{N}$ as defined in (\ref{EQ: Definition of NL-terms in perturbation equation}).

\subsection{Freely pivoting head without basal sliding}
The boundary conditions for the conjugate linear solutions $u_{0}^{+}(\bar{s})$ are given by
\begin{align}
&\dot{u}_{0}^{+}(0) = 0, \quad \dddot{u}_{0}^{+}(1) = 0, \quad u_{0}^{+}(1) + \rmi\frac{\alpha_{\mathrm{c}}}{\omega _{\mathrm{c}}}\ddot{u}_{0}^{+}(0) = 0  \nonumber \\
&\dddot{u}_{0}^{+}(0)-\alpha_{\mathrm{c}}(\dot{u}_{0}^{+}(0)+\ddot{u}_{0}^{+}(0)-\dot{u}_{0}^{+}(1))=0
\label{EQ: psi_conj_bc_pivot_with_bsliding}
\end{align}
which leads to
\begin{align}
\int_{0}^{1}u_{0}^{+}\mathcal{L}_{\mathrm{c}}u_{1} = 
- A(u_{0}^{+}(0)+\rmi\frac{\alpha_{\mathrm{c}}}{\omega _{\mathrm{c}}}\ddot{u}_{0}^{+}(0))
- B\dot{u}_{0}^{+}(1) - C\ddot{u}_{0}^{+}(0) \equiv Z
\label{EQ: Z equation - pivoting head}
\end{align}
where we have introduced
\begin{align*}
A =& \rho \rme^{\rmi\theta }\dot{u}_{0}(0)+v(0) \dot{u}_{0}(0)+w(0)\dot{u}_{0}^{\ast }(0) \\
B =& \frac{\rho \rme^{\rmi\theta }}{\bar{\alpha}_{\mathrm{c}}}\ddot{u}_{0}(1)+\frac{1}{\left\vert \bar{\alpha}_{\mathrm{c}}\right\vert ^{2}}
\frac{\bar{\beta}}{\bar{\alpha}_{\mathrm{c}}}\left\vert \ddot{u}_{0}(1)\right\vert ^{2}\ddot{u}_{0}(1) \\ 
C =& \frac{\rho \rme^{\rmi\theta }}{\bar{\alpha}_{\mathrm{c}}}\dot{u}_{0}(0)-\bar{\beta}\int_{0}^{1}
\Big[|u_{0}(\bar{s})-u_{0}(0)|^{2}(u_{0}(\bar{s})-u_{0}(0))\Big]\mathrm{d}\bar{s}\\ 
&+\frac{\bar{\alpha}_{\mathrm{c}}}{\rmi\bar{\omega}_{\mathrm{c}}}
\Big[\rho \rme^{\rmi\theta }\dot{u}_{0}(0)+\rmi\mu (u_{0}(0)-\frac{1}{\bar{\alpha}_{\mathrm{c}}}\dot{u}_{0}(0))+\int_{0}^{1}\mathcal{N}(|u_{0}|^{2}u_{0})\mathrm{d}\bar{s}\Big] \quad.
\end{align*}

\section{Shape changes of the nonlinear solutions}
\begin{figure}[!htbp]
\begin{center}
\includegraphics[width=1.0\columnwidth]{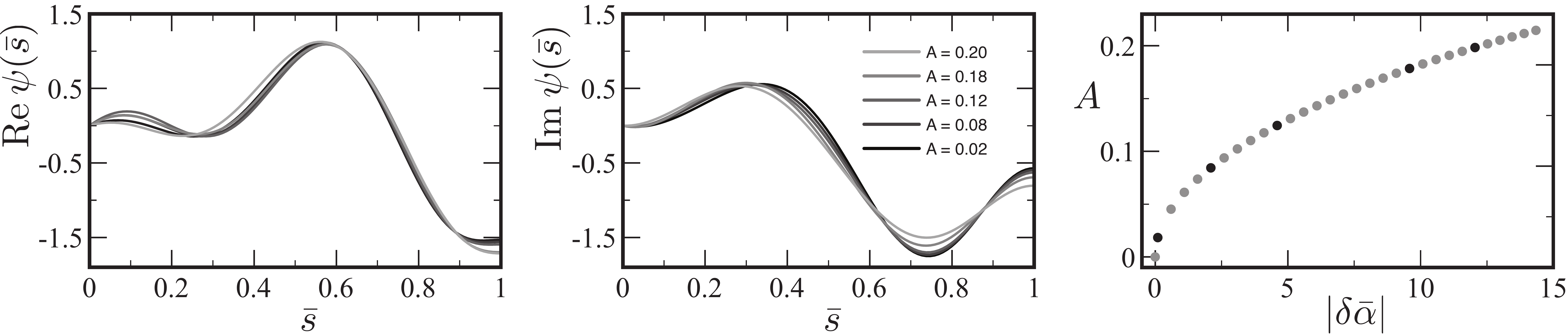}
\end{center}
\caption{Examples of the shape of the nonlinear solutions as they grow from the linearly unstable modes, corresponding to the 26 Hz solution of the second branch of unstable modes in the case of clamped head boundary conditions with basal sliding. Shown are the real and imaginary parts of the rescaled nonlinear solutions with the lighter curves corresponding to larger amplitudes. The amplitudes $A$ (in rad) of the selected modes are indicated by black dots in the diagram that shows how the amplitude grows as a function of the distance $|\delta{\bar{\alpha}}|$ to the bifurcation.}
\label{FIG: rescaled nonlinear solutions}
\end{figure}

\end{document}